\documentclass[12pt]{article}
\usepackage{epsfig}
\usepackage{latexsym}
\newcommand{\be}{\begin{equation}}
\newcommand{\ee}{\end{equation}}
\newcommand{\bea}{\begin{eqnarray}}
\newcommand{\eea}{\end{eqnarray}}

\begin{document}
\baselineskip=20pt

\vspace*{2cm}

\begin{center}
{\Large{\bf Does Inflation Provide Natural Initial 
Conditions\\ for the Universe?}\footnote{Submitted for consideration 
in the Gravity Research Foundation Essay 
Competition.  Based on {\tt hep-th/0410270}.
This work was supported in part by the
U.S. Dept. of Energy, the National Science Foundation,
the NDSEG Fellowship, and the David
and Lucile Packard Foundation.}}

\vspace*{0.3in}
Sean M.\ Carroll and Jennifer Chen
\vspace*{0.3in}

\it Enrico Fermi Institute, Department of Physics, and\\
Kavli Institute for Cosmological Physics,
University of Chicago\\
5640 S.~Ellis Avenue, Chicago, IL~60637, USA\\
{\tt carroll@theory.uchicago.edu, jennie@theory.uchicago.edu} \\
\vspace*{0.2in}
\end{center}

\begin{abstract}
If our universe underwent inflation, its entropy during the inflationary
phase was substantially lower than it is today.  Because a low-entropy 
state is less likely to be chosen randomly than a high-entropy one,
inflation is unlikely to arise through randomly-chosen initial 
conditions.  To resolve this puzzle, we examine the notion of a 
natural state for the universe, and argue that it is a nearly-empty
spacetime.  If empty space has a small vacuum energy, however, inflation
can begin spontaneously in this background.  This scenario explains
why a universe like ours is likely to have begun via a period of 
inflation, and also provides an origin for the cosmological arrow
of time.
\end{abstract}

\vfill
\newpage
\baselineskip=20pt

%%%%%%%%%%%%%%%%%%%%%%%%%%% 

As far as we know, there is only one universe.  At least, there are not
multiple universes between which we can easily travel; we only have 
access to the universe we are in, and (because there is a finite time
back to a dense early stage past which we can't see) we can only experience
a finite portion of that.

So we are stuck with the universe, and in principle we simply have to
accept the conditions we see.  Nevertheless, there is an irresistible
temptation to try to {\it explain} the state in which we find our universe as
the practically-inevitable outcome of some dynamical processes.  The most
celebrated attempt along these lines has been the theory of inflation
\cite{Guth:1980zm,Linde:1981mu,Albrecht:1982wi}, which purports to     
explain the flatness, homogeneity, and isotropy of our universe through 
an early period of accelerated expansion.

In explaining how inflation accounts for these
features of our observed universe, cosmologists often appeal to a 
certain intuitive notion of what would constitute a ``natural'' or
``randomly-chosen'' initial condition, understanding that these 
concepts are not precisely defined.  It is imagined that natural
conditions for the early universe would involve large fluctuations of
energy density and spacetime curvature, randomly distributed through
space \cite{Linde:1983gd}.  
Occasionally, a patch of space will appear 
in which the conditions are right for inflation to begin:  a smooth
configuration dominated by vacuum energy over a region larger than
the local Hubble radius \cite{Vachaspati:1998dy}.  Once inflation 
begins, that small patch can grow to the size of our observable 
universe.

However, this notion of ``natural'' cannot be right.  One way to see this is
to consider the entropy of the universe at different stages of its
evolution.  Within our current Hubble patch, the entropy due to 
particles is of order $S_M(U) \sim 10^{88}$.
This was the dominant contribution to the entropy of our comoving
Hubble patch at early times.  Today, however, the entropy is dominated 
by black holes \cite{penrose}.  
The Bekenstein-Hawking entropy of a black hole 
is proportional to its
horizon area,  $S_{BH} = A/ 4G \sim 10^{77}(M_{BH}/M_\odot)^2$.
Since there are probably more than ten billion galaxies in the 
observable universe with million-solar-mass black holes at their
centers, the current entropy in black holes is at least
$S_{BH}(U) \sim 10^{99}$.

Now let us compare this number to the entropy at the beginning of
inflation.  If inflation happens at an energy scale $M_I \sim 
10^{15}$~GeV, the Hubble radius that grew into our observable universe
was $H_I^{-1}\sim 10^6 L_P$, where $L_P\sim 10^{18}$~GeV$^{-1}$ is      
the Planck length.  The amount of entropy in this proto-inflationary
patch is approximately the entropy of an horizon volume of a de~Sitter 
spacetime with the same Hubble parameter, $S_I\sim (H_I^{-1}/L_P)^2
\sim 10^{12}$.

The entropy of the proto-inflationary patch, then, is fantastically
smaller than the entropy of our current Hubble volume, or even than
that of our comoving volume at early times before there were any
black holes.  This is in perfect accord with the Second Law of 
Thermodynamics, since the entropy is increasing.
But it is hard to reconcile with the idea that we should
find an appropriate proto-inflationary patch within  
the randomly fluctuating early universe.  If we are randomly choosing
conditions, it is much easier to choose high-entropy conditions than 
low-entropy ones; hence, it would much more likely to simply find
a patch that looks like our universe today, than to find one that
was about to begin inflating.

This point is somewhat counterintuitive, and worth emphasizing.
Despite their vast differences in size, energy, and number of particles,
the proto-inflationary patch   
and our current universe are {\it two configurations of the same
system}, since one can evolve into the other.
There are many more ways for that system to look like our
current universe than to be in a proto-inflationary configuration.
(See also \cite{penrose,Davies:1983nf,Page:1983uh,
unruh,Hollands:2002yb,Dyson:2002pf,Albrecht:2004ke}.)

Therefore, if inflation is to help explain the conditions of our 
universe, there must be some reason why it started.  To understand what
this reason might be, we should re-examine the notion of naturalness.  
Without a microscopic understanding of the statistical mechanics of quantum
gravity, we can nevertheless appeal to the fact that systems generically
evolve to high-entropy states.  In other words, we should ask what kind 
of ultimate state we expect the universe to evolve towards.

In the absence of gravity, high-entropy states tend to become 
homogeneous.  But in the presence of gravity, the Jeans instability
can lead to increased inhomogeneity, and eventually to black holes \cite{penrose}.
But a collection of black holes is not stable; spacetime
will typically be either expanding or contracting (locally), and each
black hole will eventually evaporate via Hawking radiation.  Therefore,
unless the state is arranged so that the entire universe crunches into
a future singularity, spacetime will tend towards a state which is       
nearly empty.  General relativity allows us to increase the entropy of
nearly any state by increasing the volume of space and scattering the
constituents to the far corners of the universe.

But even empty space can have energy, and recent cosmological observations
have presented strong evidence for a small nonzero vacuum energy
\cite{Riess:1998cb,Perlmutter:1998np}.  In the presence
of such an energy, empty space will settle into a de~Sitter configuration;
the natural state for the universe to be in is therefore a 
nearly-empty de~Sitter space \cite{Carroll:2004pn}.  

Unlike Minkowski space, de~Sitter has a nonzero
temperature; for realistic parameters, this temperature is $T_{\rm dS}\sim 
H_0 \sim 10^{-33}$~eV.  In the presence of
an appropriate inflaton field, thermal fluctuations will occasionally
conspire to produce a tiny,
smooth region of space dominated by a large vacuum energy -- 
the correct conditions for a proto-inflationary patch \cite{Carroll:2004pn}.  
This patch can then inflate and reheat to produce a universe like ours.  An
observer in the background would simply see the formation of a small black
hole that would eventually radiate away. (See also
\cite{Farhi:1986ty,Vilenkin:1987kf,
Farhi:1989yr,Fischler:1989se,Fischler:1990pk,Linde:1991sk}.)

Of course, thermal fluctuations in a de~Sitter background can produce all 
sorts of things, including a non-inflating Robertson-Walker universe.        
However, the fluctuations are not ``randomly chosen'' in the measure defined
by the entropy; they arise from a very specific condition, provided by
the background de~Sitter space.  Because the entropy {\it density} of
the background is so low, it is easier to fluctuate into
a small proto-inflationary patch than into a universe that looks like ours 
today.  

This story, in which inflation arises via a thermal fluctuation in a background
de~Sitter spacetime, bears a resemblance to the idea that the universe
arises as a quantum fluctuation ``from nothing'' \cite{Vilenkin:1982de,Vilenkin:1983xq,
Hartle:1983ai,Linde:1983mx,Vilenkin:1984wp}.
There is an important advantage to the current scenario, however:  it
provides a natural explanation for the cosmological arrow of time
\cite{price,albrecht03}.  The arrow-of-time puzzle can be simply stated:   
why are the initial conditions of the universe so different from the
final conditions?  In the creation-from-nothing picture, this problem
is especially acute; boundary conditions are imposed at early times but
not at late times.  

For our picture \cite{Carroll:2004pn}, in contrast, early and late times have 
the same structure
in the wider universe:  a background de~Sitter geometry that occasionally
nucleates a small inflating patch (cf. \cite{Garriga:1997ef,Aguirre:2003ck}).  
Starting from a Cauchy surface with
generic initial data, evolution to both the past and future will feature
emptying-out to a de~Sitter state, followed by eventual fluctuations
into new inflating universes.  The total entropy of the universe 
will increase because it always can increase; there  
is no equilibrium configuration in which entropy is maximized.
The mechanism for entropy increase is the creation of new inflating
patches, which can eventually evolve into universes like ours; the universe
appears time-asymmetric to us because we can only see a tiny piece of it.

We therefore believe that inflation does provide natural initial conditions
for the universe we see, once we place it in the proper context of a
larger spacetime that is stubbornly trying to increase its entropy.
It is hard to think of any directly observable consequences of this
scenario, but understanding the conceptual underpinnings of inflation
is an important part of making sense of the universe we do observe.

%%%%%%%%%%%%%%%%%%%%%%%%%%%%%%%%%%%%%%%%%%%%%%%%%%%%%

\end{document}